\documentclass[journal]{journal}
\usepackage[backend=bibtex,style=ieee,natbib=true]{biblatex}
\ifCLASSINFOpdf
\else
\fi

\usepackage[font=footnotesize]{subfig}
\usepackage{inputenc}
\usepackage{graphicx}
\graphicspath{ {./media/} }
\usepackage{multirow}

\usepackage{amsmath,amsfonts,bm}

\def\eqref#1{equation~\ref{#1}}

\def\1{\bm{1}}

\def\ve{{\bm{e}}}

\def\vg{{\bm{g}}}

\def\vr{{\bm{r}}}

\def\vv{{\bm{v}}}

\def\vx{{\bm{x}}}
\def\vy{{\bm{y}}}
\def\vz{{\bm{z}}}

\def\mB{{\bm{B}}}

\def\mE{{\bm{E}}}

\def\mR{{\bm{R}}}

\def\mX{{\bm{X}}}
\def\mY{{\bm{Y}}}
\def\mZ{{\bm{Z}}}

\DeclareMathAlphabet{\mathsfit}{\encodingdefault}{\sfdefault}{m}{sl}
\SetMathAlphabet{\mathsfit}{bold}{\encodingdefault}{\sfdefault}{bx}{n}

\newcommand{\E}{\mathbb{E}}

\usepackage [autostyle, english = american]{csquotes}
\MakeOuterQuote{"}
\usepackage{hyperref}
\usepackage{cuted}
\usepackage{flushend}
\usepackage{url}
\usepackage{amsbsy}
\usepackage{amsthm}
\usepackage{booktabs}
\usepackage{tabularx}
\usepackage{layouts}
\usepackage{pifont}
\usepackage{adjustbox}
\usepackage{xspace}

\begin{document}
\newcommand{\dropout}{\texttt{Dropout}\xspace}
\newcommand{\graphdropout}{\texttt{GraphDropout}\xspace}
\newcommand{\dense}{\texttt{Dense}\xspace}
\newcommand{\gelu}{\texttt{GELU}\xspace}
\newcommand{\relu}{\texttt{ReLU}\xspace}
\newcommand{\layernorm}{\texttt{LayerNorm}\xspace}
\newcommand{\encoder}{\texttt{Encoder}\xspace}
\newcommand{\decoder}{\texttt{Decoder}\xspace}
\newcommand{\embed}{\texttt{Embed}\xspace}
\def\xmathbf#1{\textcolor{red}{#1}}
\newcommand{\h}{\xmathbf{h}_i}
\newcommand{\x}{\xmathbf{x}}
\newcommand{\y}{\xmathbf{y}}
\newcommand{\z}{\xmathbf{z}}
\newcommand{\e}{\xmathbf{e}}

\newcommand{\m}{\xmathbf{m}_{uv}}
\newcommand{\M}{\xmathbf{m}_{uv}}
\newcommand{\curlyE}{\mathcal{E}}
\newcommand{\curlyV}{\mathcal{V}}
\newcommand{\curlyG}{\mathcal{G}}
\newcommand{\curlyX}{\mathcal{X}}
\newcommand{\curlyNu}{\mathcal{N}_u(i)}
\def\threeD{^\text{3D}}

\title{Efficient Training of Transformers for Molecule Property Prediction on Small-scale Datasets}
%
%
%

\author{Shivesh~Prakash

\thanks{S. Prakash is with the Department
of Computer Science, University of Toronto, Canada,
ON, M5S 2E4 E-mail: shivesh.prakash@mail.utoronto.ca}
}
%
%

\markboth{Journal of \LaTeX\ Class Files,~Vol.~6, No.~1, January~2007}%
{Shell \MakeLowercase{\textit{et al.}}: Bare Demo of IEEEtran.cls for Journals}
%



\maketitle
\thispagestyle{empty}

\begin{abstract}
The blood-brain barrier (BBB) serves as a protective barrier that separates the brain from the circulatory system, regulating the passage of substances into the central nervous system. Assessing the BBB permeability of potential drugs is crucial for effective drug targeting. However, traditional experimental methods for measuring BBB permeability are challenging and impractical for large-scale screening. Consequently, there is a need to develop computational approaches to predict BBB permeability. This paper proposes a GPS Transformer architecture augmented with Self Attention, designed to perform well in the low-data regime. The proposed approach achieved a state-of-the-art performance on the BBB permeability prediction task using the BBBP dataset, surpassing existing models. With a ROC-AUC of 78.8\%, the approach sets a state-of-the-art by 5.5\%. We demonstrate that standard Self Attention coupled with GPS transformer performs better than other variants of attention coupled with GPS Transformer.
\end{abstract}


%
\IEEEpeerreviewmaketitle

\section{Introduction}
The blood-brain barrier acts as a protective shield, separating the brain from the circulatory system of the body. Its primary function is to restrict the passage of solutes from the bloodstream into the central nervous system, where neurons reside~\citep{daneman2015blood}.

For drugs to effectively act on the brain, they must be capable of crossing the blood-brain barrier. Conversely, drugs intended for peripheral action should exhibit limited ability to penetrate this barrier in order to avoid undesired effects on the central nervous system. Therefore, it is crucial to determine the blood-brain barrier permeability of potential drugs. However, experimental methods for measuring brain-blood permeability are challenging, time-consuming, expensive, and impractical for large-scale screening of chemicals~\citep{thai2020fast}.

The blood-brain barrier's highly selective nature enables endothelial cells to tightly regulate the central nervous system's homeostasis~\citep{LEE2007691}. This regulation is vital for proper neuronal function and provides protection against toxins, pathogens, inflammation, injury, and disease. However, the restrictive properties of the blood-brain barrier pose a significant obstacle to drug delivery into the central nervous system. Consequently, extensive research has been conducted to develop methods for modulating or circumventing the blood-brain barrier to enhance therapeutic delivery~\citep{daneman2015blood}.

The process of brain entry for compounds is multifaceted and influenced by various factors. Lipophilic drugs, for instance, can passively diffuse across the blood-brain barrier, with their ability to form hydrogen bonds playing a role. In contrast, polar molecules typically face difficulties in crossing this barrier, although active transport mechanisms can facilitate their penetration~\citep{ABRAHAM19941257}. Local hydrophobicity, ionization profile, molecular size, lipophilicity, and flexibility are among the important parameters influencing blood-brain barrier permeation~\citep{crivori2000predicting}.

For compounds targeting the central nervous system, it is crucial to achieve brain penetration, ensuring the site of action is reached effectively~\citep{liu2004development}. Conversely, compounds designed for peripheral targets should minimize blood-brain barrier penetration to reduce potential central nervous system-related side effects. Therefore, selecting compounds with appropriate brain penetration properties is a critical consideration during the drug discovery phase~\citep{liu2004development}.

In this paper we focus on producing more powerful results using smaller datasets, thus increasing the computational efficiency.

The contributions of this paper can be summarized as:
\begin{itemize}
    \item A GPS Transformer architecture similar to GPS~\citep{rampavsek2022recipe} and GPS++~\citep{masters2022gps} paired with Self Attention which is able to perform well in the low-data regime is proposed.
    \item With this approach a new state-of-the-art for the task of predicting Blood-Brain Barrier permeability using the BBBP dataset by~\citet{wu2018moleculenet} was established, and to the best of our knowledge this is the first work to use Self Attention coupled with the GPS Transformer to improve models in this domain. A $78.8$ ROC-AUC was achieved, beating the current state-of-the-art~\citep{chang2023bidirectional} by $5.5$.
\end{itemize}

\section{Related Work}
The paper by~\citet{miao2019improved} proposes a Deep Learning method for predicting the blood-brain barrier permeability based on clinical phenotypes data. The method aims to overcome the limitations of existing prediction approaches that rely on physical characteristics and chemical structure of drugs, which are typically applicable only to small molecule compounds that passively diffuse through the BBB.

The Deep Learning method leverages clinical phenotypes data to train a predictive model. Clinical phenotypes refer to observable characteristics or traits associated with drugs, such as drug side effects and drug indications. By incorporating this information into the prediction model, the method aims to capture more comprehensive and complex mechanisms of BBB penetration beyond passive diffusion.

\citet{miao2019improved} validate their Deep Learning method using three datasets. The validation results demonstrate that their approach outperforms existing methods in terms of prediction accuracy. Specifically, the average accuracy achieved by their method is 0.97, the area under the receiver operating characteristic curve (AUC) is 0.98, and the F1 score is 0.92. These performance metrics indicate a high level of accuracy in predicting BBB permeability. By incorporating clinical phenotypes data into the predictive model, the method demonstrates superior performance compared to existing physical, chemical, and supervised learning approaches~\citep{miao2019improved}.

\citet{jain2022silico} in their paper address the limitations of existing models used to predict blood-brain barrier (BBB) permeability by exploring the role of inflammation in influencing BBB permeability. Inflammation, measured through acute phase C-reactive protein (CRP) levels, has been found to have a direct correlation with BBB permeability. A threshold of 2.5 $\mu$g/ml of CRP is established, above which BBB impairment is expected.

To overcome the challenges of existing models, a custom-built dataset of 281 molecules is created, and a machine learning approach is employed. Initially, different models such as multi-layer perceptron regression (MPR) and ensemble models are tested, but they struggle with outliers and lack the ability to predict descriptors for future drugs. Finally, a fully connected neural network (FCNN) is chosen as the predictive logBB model, which performs well in terms of learning the distribution and achieving lower error levels.

The research also incorporates a neuroinflammation model, which considers CRP levels alongside the logBB model. Although CRP levels do not directly correlate with logBB values, the model explores the second and third-order feature derivatives that show significant correlations. A quadratic polynomial regression model is used to determine this correlation. The software developed for this research includes preprocessing, the predictive logBB model, and the neuroinflammation model, providing a continuous logBB value that represents BBB permeability based on the patient's inflammation level. This approach aims to improve the accuracy of predicting BBB permeability for drugs and better understand the impact of inflammation on BBB function.

The study by~\citet{saber2020prediction} focuses on predicting drug permeability across the blood-brain barrier (BBB) using in silico models. The researchers compare the performance of sequential feature selection (SFS) and genetic algorithms (GA) in selecting relevant molecular descriptors to enhance the accuracy of permeability prediction. Five different classifiers are trained initially using eight molecular descriptors, and then SFS and GA are separately applied to choose the descriptors for each algorithm.

The results show that both SFS and GA improve the accuracy of the classifiers, but GA outperforms SFS. The highest accuracy of 96.23\% is achieved with GA, specifically with a fitness function based on the performance of a support vector machine. The study highlights the significance of the polar surface area (PSA) of drugs in crossing the BBB. GA consistently selects the PSA and the number of hydrogen bond donors as the most relevant descriptors, providing better results compared to using other features.

The findings suggest that GA is a more robust approach for selecting relevant descriptors in predicting BBB permeability. The selected classifiers demonstrate a good balance in predicting BBB+ and BBB- drugs. Accurate in silico BBB models are crucial for early-phase drug discovery, reducing the need for extensive in vitro testing and saving time and resources.

The paper by~\citet{yuan2018fingerprint} describes the importance of predicting the permeability of compounds through the blood-brain barrier (BBB) for drug discovery targeting the brain. It discusses the use of computational methods, particularly support vector machine (SVM), in predicting BBB permeability. Different types of descriptors, such as molecular property-based descriptors (1D, 2D, and 3D descriptors) and fragment-based descriptors (fingerprints), have been utilized in SVM models. The selection of descriptors greatly influences the performance of the SVM model. 

The paper aims to develop a new SVM model by combining molecular property-based descriptors and fingerprints to improve the accuracy of BBB permeability prediction. The results indicate that the proposed SVM model outperforms existing models in predicting BBB permeability. The paper also provides an overview of the blood-brain barrier and the complex factors that influence compound penetration into the brain. Additionally, it discusses the classification of quantitative and qualitative BBB permeability prediction models and highlights SVM as a superior method for qualitative prediction. The paper concludes that the combined use of property-based descriptors and fingerprints improves the accuracy of SVM models and suggests that similar approaches may enhance computational predictions for other molecular activities in the future.

However, our work explores how we find a simple modification to a General, Powerful, Scalable Graph Transformer (GPS)~\citet{rampavsek2022recipe} and improve upon the performance for the task of blood-brain barrier permeability while showing the effectiveness of simple methods within the graph setting.

\section{Data}

We perform our experiments on the dataset 'BBBP: Binary labels of blood-brain barrier penetration (permeability)' which is a classification dataset available as a part of MoleculeNet~\citep{wu2018moleculenet} which contains 2,050 molecules and each molecule comes with a name, label, and SMILES string. The label is a boolean integer which denotes if a given compound can pass through the blood-brain barrier or not. 

The Blood–brain barrier penetration (BBBP) dataset comes from a study by~\citet{martins2012bayesian} on the modeling and prediction of the barrier permeability using a Bayesian approach. For this well-defined target scaffold splitting is also recommended by~\citet{wu2018moleculenet}.

In the paper by~\citet{martins2012bayesian}, a unique approach based on Bayesian statistics, in conjunction with state-of-the-art machine learning techniques, was employed to create a robust model suitable for real-world drug research applications. The dataset used for the Bayesian analysis consisted of 1970 carefully curated molecules, making it one of the largest, but still small, datasets used in similar studies. Various configurations of Random Forests and Support Vector Machines, coupled with different combinations of chemical descriptors, were evaluated. To assess the performance of the models, a 5-fold cross-validation process was employed, and the best-performing model was further tested on an independent validation set.

The model proposed by~\citet{martins2012bayesian} achieved an impressive overall accuracy of 95\%, as measured by a mean square contingency coefficient ($\phi$) of 0.74. Furthermore, this model exhibited a high capacity for predicting blood-brain barrier (BBB) positives, with an accuracy of 83\%, and a remarkable accuracy of 96\% in determining BBB negatives. These findings highlight the efficacy of the developed model in accurately predicting the permeability of molecules across the blood-brain barrier, based on experiments by~\citet{martins2012bayesian}.

\section{Methods}

The proposed approach closely follows the framework of General, Powerful, Scalable Graph Transformer (GPS) by~\citet{rampavsek2022recipe} and that of GPS++ by~\citet{masters2022gps}. This paper explores how these frameworks could be used to learn from low amounts of data. Thus, the BBBP dataset was used which contains only 2050 molecules.

\subsection{GPS Block}

\newcommand{\mpnn}{\texttt{MPNN \xspace}}
\newcommand{\attn}{\texttt{GlobalAttn \xspace}}
\newcommand{\X}{\mathbf{X}}
\newcommand{\A}{\mathbf{A}}
\citet{masters2022gps} proposed that at each layer, the features are updated by aggregating the output of an MPNN layer with that of a global attention layer which is described by the equations below. Note that the edge features are only passed to the \mpnn layer, and that residual connections with batch normalization~\cite{ioffe2015batchnorm} are omitted for clarity. Both the \mpnn and \attn layers are modular, i.e., \mpnn can be any function that acts on a local neighborhood and \attn can be any fully-connected layer.
\begin{align}
    \X^{\ell+1}, \E^{\ell+1} &= \texttt{GPS}^{\ell} \left( \X^{\ell}, \E^{\ell}, \A \right),\\
    \textrm{computed as} \ \ \ \ 
    \X^{\ell+1}_M, \ \E^{\ell+1} &= \mpnn_e^{\ell} \left(\X^{\ell}, \E^{\ell}, \A \right),\\
    \X^{\ell+1}_T &=
    \attn^{\ell} \left(\X^{\ell}  \right),\\
    \X^{\ell+1} &=
    \texttt{MLP}^{\ell}\left(\X^{\ell+1}_M + \X^{\ell+1}_T\right)
    \label{eqn:layer_equation}
\end{align}
where $\A \in \mathbb{R}^{N \times N}$ is the adjacency matrix of a graph with $N$ nodes and $E$ edges; $\X^{\ell} \in \mathbb{R}^{N \times d_\ell}, \E^{\ell} \in \mathbb{R}^{E \times d_\ell}$ are the $d_\ell$-dimensional node and edge features, respectively; $\mpnn_e^{\ell}$ and $\attn^{\ell}$ are instances of an MPNN with edge features and of a global attention mechanism at the $\ell$-th layer with their corresponding learnable parameters, respectively; $\texttt{MLP}^{\ell}$ is a 2-layer MLP block.

\subsection{GPS++ Block}

\def\vy{\mathbf{y}}
\def\mY{\mathbf{Y}}
\def\vz{\mathbf{z}}
\def\mZ{\mathbf{Z}}
\citet{rampavsek2022recipe} proposed the GPS++ block to be defined as follows for layers $\ell > 0$:
\def\l{^\ell}
\def\xin{\mathbbm{x}}
\def\ein{\mathbbm{e}}
\def\mR{\mathbf{R}}
\def\vr{\mathbf{r}}
\def\vx{\mathbf{x}}
\def\mX{\mathbf{X}}
\def\ve{\mathbf{e}}
\def\mE{\mathbf{E}}
\def\vg{\mathbf{g}}
\def\xdim{{d_\text{node}}}
\def\edim{{d_\text{edge}}}
\def\myforall #1:{\forall #1:~~~}
\def\gdim{{d_\text{global}}}
\newcommand{\gps}{\texttt{GPS++}}
\newcommand{\ffn}{\texttt{FFN}}
\begin{align}\label{eqn:gps_layer}
    \mX^{\ell+1}, \mE^{\ell+1}, \vg^{\ell+1} &= \gps \left( \mX\l, \mE\l, \vg\l, \mB \right) \\
    \mY\l, \ \mE^{\ell+1}, \ \vg^{\ell+1} &= \mpnn \left(\mX\l, \mE\l, \vg\l \right),\\
    \mZ\l &=  \texttt{BiasedAttn} \left(\mX\l, \mB  \right),\\
    \myforall i: \vx_i^{\ell+1} &= \ffn\left(\vy_i\l + \vz_i\l\right)
\end{align}

In \gps~\citep{masters2022gps} which is analogous to \mpnn modules as shown by~\citet{gilmer2017neural, battaglia2018relational}, and~\citet{bronstein2021geometric}. They indicate that their \mpnn variant maximizes the expressivity of the model and increases the generalizability of the model.

\subsection{Our Architecture}

The proposed block is defined as follows which is similar to the construction of \gps~\citep{masters2022gps}.

\def\RR#1{\mathbb{R}^{#1}}
\def\brak#1{\bigl[#1\bigr]}
\def\hcat{~\mid~} \def\hcatname{a vertical bar}
\def\Bigghcat{~~~\Bigg|~}
\def\vcat{~;~} \def\vcatname{a semicolon}
\def\vfor{~\mathsf{for}~}
\def\vv{\mathbf v}
\begin{align}\label{eqn:our_layer}
    \mX^{\ell+1}, \mE^{\ell+1}, \vg^{\ell+1} &= \gps \left( \mX\l, \mE\l, \vg\l, \mB \right) \\
    \mY\l, \ \mE^{\ell+1}, \ \vg^{\ell+1} &= \mpnn \left(\mX\l, \mE\l, \vg\l \right),\\
    \mZ\l &=  \texttt{Attn} \left(\mX\l, \mB  \right),\\
    \myforall i: \vx_i^{\ell+1} &= \ffn\left(\vy_i\l + \vz_i\l\right)
\end{align}

where $\mX^{0}$, $\mE^{0}$, $\vg^0$, and $\mB$ defined in Equation \ref{eq:inits}.

\begin{figure*}[!t]
\normalsize
\begin{align}
\label{eq:inits}
    \mX^{0}&=\dense(\brak{\mX^\text{atom} \hcat \mX^\text{LapVec} \hcat \mX^\text{LapVal} \hcat \mX^\text{RW} \hcat \mX^\text{Cent} \hcat \mX\threeD}) &&\in\RR{N\times \xdim} \nonumber\\
    \mE^{0}&= \dense(\brak{\mE^\text{bond} \hcat \mE\threeD}) &&\in\RR{M\times \edim}\\
    \vg^0 &= \embed_{\gdim}(0) &&\in\RR{\gdim} \nonumber\\
    \mB&=\mB^{\mathrm{SPD}} + \mB\threeD &&\in\RR{N\times N}\nonumber
\end{align}
\end{figure*}

\begin{figure}
    \centering
    \includegraphics[width=\columnwidth]{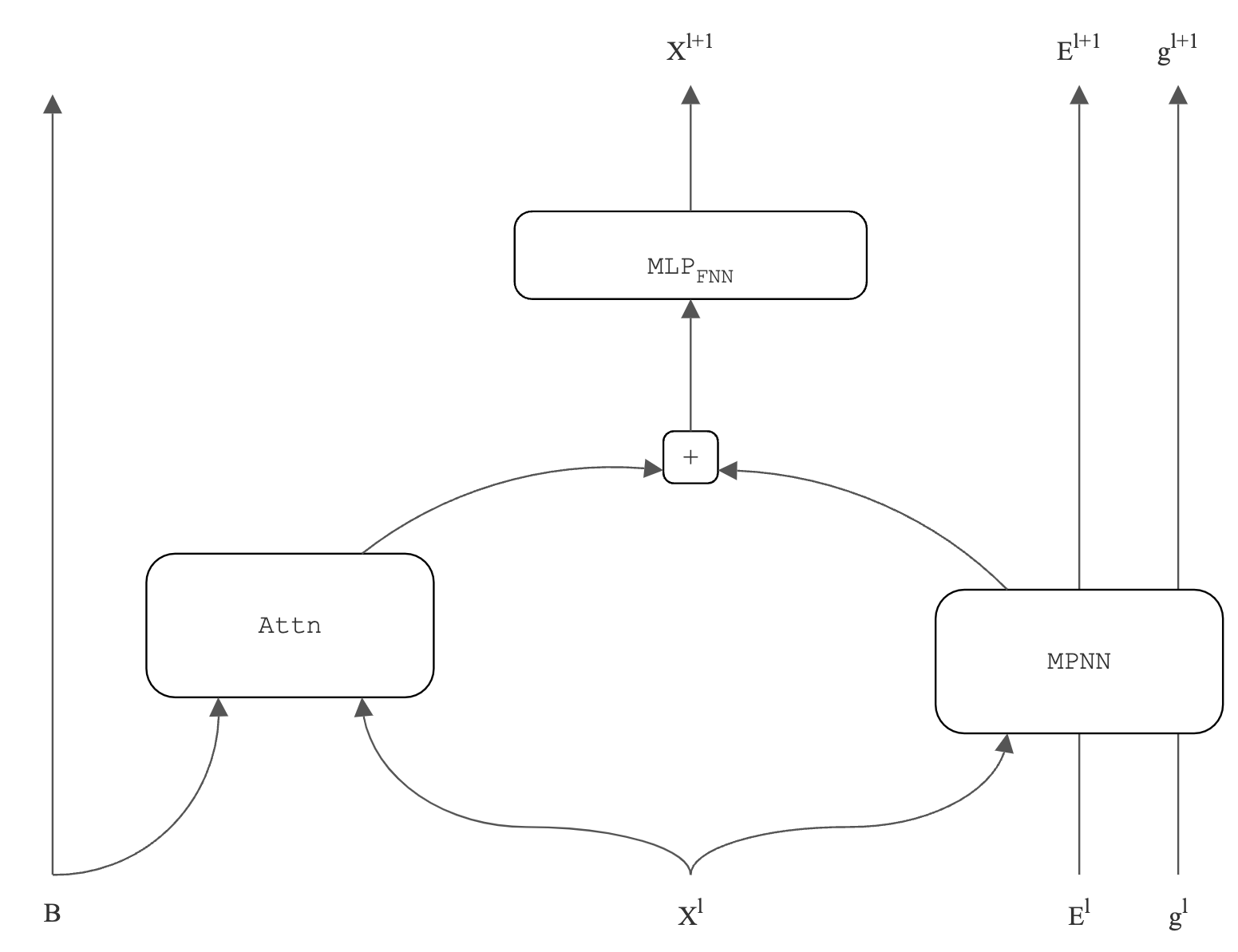}
    \caption{Overview of the proposed model, notice the architectural changes made from GPS++.}
    \label{fig:architecture}
\end{figure}

{We} do not use the MPNN variant proposed by GPS++~\citep{masters2022gps} in {our} architecture which increases the overfitting considering the sizes of the datasets {we} work with. {Our} approach also incorporates \texttt{Attn} which is defined as the standard attention block from~\citet{vaswani2017attention}. {We} also find that for smaller graph datasets standard Attention, modules work better than Attention variants like Biased self-attention~\citep{NEURIPS2021_f1c15925} which was used by GPS++~\citep{masters2022gps}. The Feed Forward Network is defined simply as a stack of MLP layers and also uses Drop GNN~\citep{papp2021dropgnn}. These architectural choices are also summarized in Figure~\ref{fig:architecture}.

With these changes in the Architecture of GPS++, our approach outperforms GPS++ on smaller datasets and sets a new state-of-the-art on the BBBP dataset achieving $78.8$ ROC-AUC, beating the current state-of-the-art~\citep{chang2023bidirectional} by $5.5$.

\section{Experimental Setup}

This section explains the implementation details of the experiments and \text{our} proposed model.

\subsection{Sampling}

There is a class imbalance in the dataset, meaning that not all classes have a similar number of images. For this reason, {we} follow a stratified sampling strategy during data loading to ensure each batch contains $50 \pm 5\%$ instances of each label class.

\subsection{Model backbone}

Throughout this work GPS~\citep{rampavsek2022recipe} and GPS++~\citep{masters2022gps} is used as the architecture backbone due to their success in using MPNNs and Transformers. {Our} model backbone is standard self-attention~\citep{vaswani2017attention} which {we} demonstrate works well for the task of molecule property prediction. Since the BBBP dataset does not contain a large amount of data, other transformer-based models did not show great results for this task whereas standard self-attention model with regularization and augmentation techniques, was able to generalize well and achieved better results.

\subsection{Baseline model}

{We} established a naive baseline with random guesses. The baseline model {we} chose was training GPS++~\citep{masters2022gps} without any design modifications. This gets to a ROC-AUC of $72.7$ on the BBBP dataset. To train this model, {we} employ standard augmentations using AugLiChem~\citep{Magar_2022} and train the network for 300 epochs with a batch size of 256.

\subsection{Loss function}

The final model prediction is formed by global sum-pooling of all node representations and then passing it through a MLP. The regression loss is the mean absolute error (L1 loss) between a scalar prediction and the ground truth.

\subsection{Code}

{Our} code is in PyTorch 1.10~\citep{paszke2019pytorch} and we also provide TensorFlow code. We use a number of open-source packages to develop {our} training workflows. Most of {our} experiments and models were trained with PyTorch Geometric~\citep{Fey_Fast_Graph_Representation_2019}. {Our} hardware setup for the experiments included either four NVIDIA Tesla V100 GPUs or a TPUv3-8 cluster. {We} utilized mixed-precision training with PyTorch's native AMP (through \texttt{torch.cuda.amp}) for mixed-precision training and a distributed training setup (through \texttt{torch.distributed.launch}) which allowed {us} to obtain significant boosts in the overall model training time. Our code to reproduce the results along with Tensorboard logs of the training runs are available at \footnote{\url{https://github.com/Rishit-dagli/GraphBRAIN}}.

\section{Results}

{We} report all results and compare them against previous models and a random baseline (equivalent to making a guess) in Table~\ref{tab:results}. The performance of models is calculated using the metrics that are typical for a molecular property prediction problem, ROC-AUC. Additionally, it is found that using any of the other variants of GPS++ is detrimental for this task and the model starts heavily overfitting even with {our} design changes, augmentation, and regularization. {We} also observe that {our} architectural changes over GPS++~\citep{masters2022gps} improves GPS++ by $6.1\%$.

\begin{table}[ht]
    \centering
    \caption{Model Performance on the BBBP dataset for predicting blood-brain membrane permeability.}
    \begin{adjustbox}{width=\columnwidth,center}
    \begin{tabular}{lr}
    \toprule
    \textbf{Method Description} & \textbf{ROC-AUC ($\uparrow$)}\\
    \midrule
    GAL 120B~\citep{taylor2022galactica} & 66.1\\
    PretrainGNN~\citep{hu2020strategies} & 68.7\\
    N-GramRF~\citep{liu2019ngram} & 69.7\\
    GROVER~\citep{rong2020selfsupervised} & 70.0\\
    D-MPNN~\citep{yang2019analyzing} & 71.0\\
    ChemRL-GEM~\citep{Fang_2022} & 72.4\\
    ChemBERTa-2~\citep{ahmad2022chemberta2} & 72.8\\
    Uni-Mol~\citep{taylor2022galactica} & 72.9\\
    SPMM~\citep{chang2023bidirectional} & 73.3\\
    \midrule
    GPS & 72.34\\
    GPS++ & 72.73\\
    \textbf{Ours (GPS Transformer with Self Attention)} & \textbf{78.80}\\
    \bottomrule
    \end{tabular}
    \end{adjustbox}
    \label{tab:results}
\end{table}

\section{Conclusion}

This research has provided valuable insights into the challenges and opportunities associated with drug delivery across the blood-brain barrier. By understanding the intricate mechanisms of brain entry and the parameters influencing blood-brain barrier permeation, this paper have paved the way for the development of novel strategies in the field of therapeutic delivery. The introduction of GPS Transformer architecture, combined with Self Attention, has demonstrated significant advancements in predicting blood-brain barrier permeability using limited data. Notably, the approach has surpassed the current state-of-the-art performance, as evidenced by achieving an impressive ROC-AUC of 78.8. This breakthrough opens up new possibilities for efficient and effective drug discovery and development, ultimately aiming to improve treatments for central nervous system disorders. The findings contribute to the bigger vision of enhancing patient care by enabling targeted drug delivery to the brain, thus revolutionizing common practices and offering hope for transformative advancements in neurological medicine.

\section*{Data Availability}

BBBP data was obtained from MoleculeNet~\citep{wu2018moleculenet} and is available at the following URL: \hyperlink{https://moleculenet.org/datasets-1}{https://moleculenet.org/datasets-1}.

\printbibliography

%







\end{document}